\documentclass[a4paper,10pt]{article}
\usepackage[T2A]{fontenc}
\usepackage{amssymb,amsfonts,amsmath,mathtext,cite,enumerate,float,mathrsfs}
\usepackage[utf8]{inputenc}
\usepackage[english]{babel}
\usepackage{graphicx,color}
\usepackage{wrapfig}
\graphicspath{{./}}

\usepackage{epsfig}
\usepackage{epsf}
\usepackage{subfig} 
\usepackage{feynmp} 
\usepackage{pgf,pgfarrows,pgfnodes,pgfautomata,pgfheaps,multirow}
\usepackage{array}
\usepackage{bm}
\usepackage{latexsym}
\usepackage{pifont}

\newcommand{\be}{\begin{equation}}
\newcommand{\ee}{\end{equation}}
\newcommand{\bea}{\begin{eqnarray}}
\newcommand{\eea}{\end{eqnarray}}
\newcommand\nn{\nonumber}

\title{The processes $e^{+}e^{-} \to \pi\pi(\pi')$ in the extended NJL model}

\author{M. K. Volkov\footnote{E-mail address: volkov@theor.jinr.ru}, D. G. Kostunin\footnote{E-mail address: kostunin@theor.jinr.ru}\\
\it Bogoliubov Laboratory of Theoretical Physics, JINR\\ 
\it Dubna, 141980, Russia}

\begin{document}

\maketitle

\begin{abstract}
The process $e^{+}e^{-} \to \pi^{+}\pi^{-}$ is described in the framework of the extended NJL model. Intermediate vector mesons $\rho^0(770)$, $\omega(782)$ and $\rho'(1450)$ are taken into account.
Our results are in satisfactory agreement with experimental data.
The prediction for the process $e^{+}e^{-} \to \pi\pi'(1300)$ is given. Here the main contribution is given by the diagram with intermediate $\rho'(1450)$ meson.
\\

\\

\end{abstract}



\section{Introduction}
Recently, the processes $e^{+}e^{-} \to \pi^0 \gamma,\,\, \pi' \gamma,\,\, \pi^0 \omega,\,\, \pi^0 \rho^0\,, \pi^0 \pi^0 \gamma$ have been described in the framework of the extended NJL model~\cite{Arbuzov:2010xi,Ahmadov:2011qv,ArbuzovKuraevVolkov,Ahmadov:2011ey}.
Intermediate vector mesons in the ground and excited states were took into account.
In all these works satisfactory agreement with experimental data was obtained without using any arbitrary parameters.

In this work the same method is used for description of the processes $e^{+}e^{-} \to \pi\pi(\pi')$.
Now this process is thoroughly investigated from both experimental~(see \cite{SND} and other references in this work) and theoretical~\cite{Gounaris:1968mw,kuhn,O'Connell:1995wf,Dominguez:2007dm,Jegerlehner:2011ti,Achasov:2011ra} points of view.
However, in all these works it is necessary to use the number of additional parameters.
The extended NJL model allows us to describe these processes without attraction of additional parameters.

Let us note that for the description of the processes $e^{+}e^{-} \to \pi^{+}\pi^{-}$ with intermediate photon and vector mesons in the ground state it is sufficient to use the standard NJL model~\cite{VolkovEbert,VolkovAn,pepan86,EbertReinhardt,pepan93,VolkovEbertReinhardt,HatsudaKunihiro,UFN,Vogl:1991qt}.
On the other hand, for the description of the amplitude of this process with the intermediate $\rho'(1450)$ meson the extended NJL model~\cite{VolkovWeiss,yadPh,VolkovEbertNagy,VolkovYudichev,UFN} should be used.

\section{Lagrangian and process amplitudes}

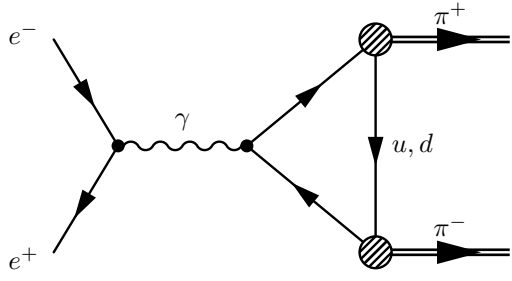
\begin{figure}[h]
\begin{center}

\begin{fmffile}{contact}

      \begin{fmfgraph*}(200,100)
	      \fmfpen{thin}\fmfleftn{l}{2}\fmfrightn{r}{2}

	      \fmfright{b,a}
	      \fmfleft{f,fb}
	      \fmflabel{$e^{-}$}{fb}
	      \fmflabel{$e^{+}$}{f}
	      \fmf{fermion}{fb,v1,f}
 	      \fmf{fermion,tension=.5}{p1,v2,p2}
	      \fmf{boson,lab.side=left,label=$\gamma$}{v1,v2}
 	      \fmf{fermion,lab.side=left,label=$u,,d$}{p2,p1}	      
 	      \fmfdotn{v}{2}
	      \fmfblob{0.06w}{p2}
	      \fmfblob{0.06w}{p1}

	      \fmf{dbl_plain_arrow,lab.side=left,label=$\pi^{+}$}{p2,a}
	      \fmf{dbl_plain_arrow,lab.side=left,label=$\pi^{-}$}{p1,b}

	      \fmfforce{120,90}{p2}
	      \fmfforce{120,10}{p1}

	      \fmfforce{170,90}{a}
	      \fmfforce{170,10}{b}

	      \fmfforce{0,90}{fb}
	      \fmfforce{0,10}{f}

      \end{fmfgraph*}

\end{fmffile}
\caption{Contact interaction of the photon with a pion pair through the photon propagator}
\label{fig1}
\end{center}
\end{figure}

\begin{figure}[h]
\begin{center}

\begin{fmffile}{rho}

      \begin{fmfgraph*}(200,100)
	      \fmfpen{thin}\fmfleftn{l}{2}\fmfrightn{r}{2}

	      \fmfright{b,a}
	      \fmfleft{f,fb}
	      \fmflabel{$e^{-}$}{fb}
	      \fmflabel{$e^{+}$}{f}
	      \fmf{dbl_plain_arrow,lab.side=left,label=$\rho,,\omega,,\rho^{'}$}{v1,v4}
	      \fmf{boson,label=$\gamma$}{v2,v3}
	      \fmf{fermion,left,tension=.5}{v1,v2}
	      \fmf{fermion,label=$u,,d$,left,tension=.5}{v2,v1}
	      \fmf{fermion}{fb,v3,f}
 	      \fmf{fermion,tension=0.5}{p1,v4,p2}
 	      \fmf{fermion,lab.side=left,label=$u,,d$}{p2,p1}	      
	      \fmfdotn{v}{4}
	      \fmfblob{0.06w}{p2}
	      \fmfblob{0.06w}{p1}

	      \fmf{dbl_plain_arrow,lab.side=left,label=$\pi^{+}$}{p2,a}
 	      \fmf{dbl_plain_arrow,lab.side=left,label=$\pi^{-}$}{p1,b}

	      \fmfforce{220,90}{p2}
	      \fmfforce{220,10}{p1}

	      \fmfforce{270,90}{a}
	      \fmfforce{270,10}{b}

	      \fmfforce{0,90}{fb}
	      \fmfforce{0,10}{f}

      \end{fmfgraph*}

\end{fmffile}
\caption{Interaction with intermediate vector mesons}
\label{fig2}
\end{center}
\end{figure}
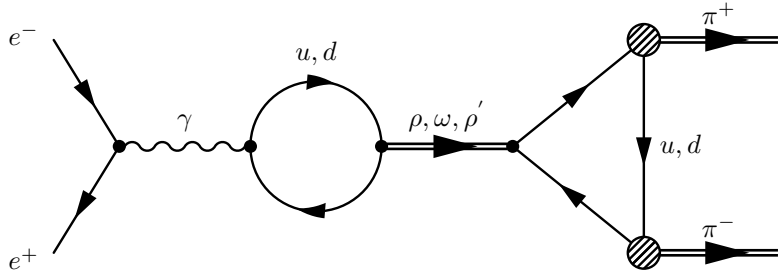

The amplitude of the process $e^{+}e^{-} \to \pi^{+}\pi^{-}$ is described by the diagrams given on Figs.~\ref{fig1} and~\ref{fig2}.

For the description of the first three diagrams with intermediate $\gamma$, $\rho$, and $\omega$ mesons in the ground state
we need the part of the standard NJL Lagrangian which describes interactions
of photons, pions, and the ground states of vector mesons with quarks

\bea \label{L1}
&& \Delta{\mathcal L}_{1} = \bar{q}\biggl[i\hat\partial - m
+ \frac{e}{2}\biggl(\tau_3+\frac{1}{3}I\biggr)\hat{A}
+ig_{\pi}\gamma_5\tau_{\pm}\pi^{\pm}
+ \frac{g_\rho}{2}\gamma_\mu\left(I\hat{\omega} + \tau_3\hat{\rho}^0\right) \biggr]q,
\eea
where $\bar{q}=(\bar{u},\bar{d})$ with $u$ and $d$ quark fields;
$m=diag(m_u,m_d)$, $m_u=280$~MeV is the constituent quark mass, $m_d - m_u \approx 3.7$~MeV as will be shown below;
$e$ is the electron charge;
$A$, $\pi^{\pm}$, $\omega$ and $\rho^0$ are the photon, pion, $\omega$ and $\rho$ meson fields, respectively;
$g_\pi$ is the pion coupling constant,
$g_\pi=m_u/f_\pi$, where $f_\pi=93$~MeV is the pion decay constant;
$g_\rho$ is the vector meson coupling constant, $g_\rho\approx 6.14$ corresponds
to the standard relation $g_\rho^2/(4\pi)\approx 3$;
$\tau_{\pm} = (\tau_1 \mp i\tau_2)/\sqrt{2}$,
$I=diag(1,1)$ and $\tau_{1,2,3}$ are Pauli matrices.

For description of the radial excited mesons interactions we use the extended version
of the NJL Lagrangian~\cite{yadPh,VolkovEbertNagy,ArbuzovKuraevVolkov}:
\bea \label{L2}
&& \Delta {\mathcal L}_2^{\mathrm{int}} =
\bar{q}(k')\biggl\{ A_\pi \tau_{\pm}\gamma_5\pi(p)
- A_{\pi'} \gamma_5\tau_{\pm}\pi'(p)
+ A_{\omega,\rho} \left(\tau_3{\hat{\rho}}(p)+I\hat{\omega}(p)\right)
- A_{\rho'} \tau_3{\hat{\rho}'}(p)
\biggr\}q(k),
 \\ \nn   
&& p = k-k',
 \\ \nn
&& A_\pi = g_{\pi_1}\frac{\sin(\alpha+\alpha_0)}{\sin(2\alpha_0)}
       +g_{\pi_2}f({k^\bot}^2)\frac{\sin(\alpha-\alpha_0)}{\sin(2\alpha_0)},
\nn \\
&& A_{\pi'} = g_{\pi_1}\frac{\cos(\alpha+\alpha_0)}{\sin(2\alpha_0)}
       +g_{\pi_2}f({k^\bot}^2)\frac{\cos(\alpha-\alpha_0)}{\sin(2\alpha_0)},
\nn \\
&& A_{\omega,\rho} = g_{\rho_1}\frac{\sin(\beta+\beta_0)}{\sin(2\beta_0)}
       +g_{\rho_2}f({k^\bot}^2)\frac{\sin(\beta-\beta_0)}{\sin(2\beta_0)},
\nn \\ \nn
&& A_{\rho'} = g_{\rho_1}\frac{\cos(\beta+\beta_0)}{\sin(2\beta_0)}
        +g_{\rho_2}f({k^\bot}^2)\frac{\cos(\beta-\beta_0)}{\sin(2\beta_0)}.
\eea
The radially-excited states were introduced in the NJL model with the help of
the form factor in the quark-meson interaction:
\bea
f({k^\bot}^2) &=& (1-d |{k^\bot}^2|) \Theta(\Lambda^2-|{k^\bot}^2|),
\nn \\
{k^\bot} &=& k - \frac{(kp)p}{p^2},\qquad d = 1.788\ {\mathrm{GeV}}^{-2},
\eea
where $k$ and $p$ are the quark and meson momenta, respectively.
The cut-off parameter $\Lambda=1.03$~GeV.
The coupling constants $g_{\rho_1}=g_\rho$ and $g_{\pi_1}=g_\pi$ are the same as in
the standard NJL version. The constants $g_{\rho_2}=10.56$ and $g_{\pi_2} = g_{\rho_2}/\sqrt{6}$, and the mixing
angles $\alpha_0=58.39^\circ$, $\alpha=58.70^\circ$, $\beta_0=61.44^\circ$, and $\beta=79.85^\circ$
were defined in refs.~\cite{VolkovEbertNagy,ArbuzovKuraevVolkov}.

The amplitude $e^{+}e^{-} \to \pi^{+}\pi^{-}$ has the form
\begin{equation}
T = \bar{e}\gamma_\mu e \frac{4\pi\alpha_e}{s} \left(B_{\gamma\rho} + B_\omega + B_{\rho'}\right) f_{a_1}(s) (p_{\pi^{+}}^\mu - p_{\pi^{-}}^\mu) \pi^{+} \pi^{-}\, ,
\end{equation}
where $\alpha_e = e^2/4\pi \approx 1/137$, $s = (p_{e^{+}} + p_{e^{-}})^2$, and

\begin{equation}
f_{a_1}(p^2) = Z + (1 - Z) + \left(\frac{p^2 - m_\pi^2}{(g_\rho F_\pi)^2}\right)\left(1 - \frac{1}{Z}\right) = 1 + \left(\frac{p^2 - m_\pi^2}{(g_\rho F_\pi)^2}\right)\left(1 - \frac{1}{Z}\right)\, ,
\end{equation}
where $Z = (1 - 6m_u^2/m_{a_1}^2)^{-1}$ is the additional renormalizing factor pion fields that appeared after the inclusion of $a_1$ -- $\pi$  transitions.
This function describes the creation of pions at the ends of the triangle quark diagram with taking into account the possibility of creation of these pions through the intermediate axial-vector $a_1(1260)$ meson.
The first term of this amplitude corresponds to the triangle diagram without $a_1$ -- $\pi$ transitions, the second term corresponds to diagram with $a_1$ -- $\pi$ transition on the one of the pion lines and the third term corresponds to the diagram with transitions on both pion lines\footnote{Let us note that $f_{a_1}(p^2)$ has the value which is close to the factor introduced in~\cite{Achasov:2011ra}.}.

 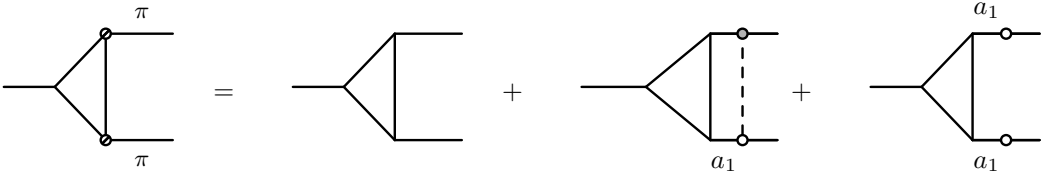
\begin{figure}[h]
 \begin{center}
 \begin{tabular}{ccccccc}

	\begin{fmffile}{pi0}
      \begin{fmfgraph*}(63,48)

	      \fmfpen{thin}\fmfleftn{l}{2}\fmfrightn{r}{2}
 
	      \fmfright{b,a}
	      \fmfleft{f,fb}
 	      \fmf{vanilla}{fb,v1}
	      \fmf{vanilla,tension=.5}{p1,v1,p2}
	      \fmf{vanilla}{p2,p1}
	      \fmfblob{0.06w}{p2}
	      \fmfblob{0.06w}{p1}
	      \fmf{vanilla,lab.side=left,label= \begin{small} $\pi$ \end{small}}{p2,a}
	      \fmf{vanilla,lab.side=down,label=\begin{small}$\pi$\end{small}}{p1,b}

	      \fmfforce{38,43}{p2}
	      \fmfforce{38,3}{p1}

	      \fmfforce{63,43}{a}
	      \fmfforce{63,3}{b}

	      \fmfforce{0,23}{fb}

      \end{fmfgraph*}  
	\end{fmffile}

&  \multirow{1}[4]{*}[5ex]{$ = $ } 

&	\begin{fmffile}{pi1}
      \begin{fmfgraph*}(63,48)

	      \fmfpen{thin}\fmfleftn{l}{2}\fmfrightn{r}{2}
 
	      \fmfright{b,a}
	      \fmfleft{f,fb}
 	      \fmf{vanilla}{fb,v1}
	      \fmf{vanilla,tension=.5}{p1,v1,p2}
			\fmf{vanilla}{p2,p1}
	      \fmf{vanilla}{p2,a}
	      \fmf{vanilla}{p1,b}
   
	      \fmfforce{38,43}{p2}
	      \fmfforce{38,3}{p1}

	      \fmfforce{63,43}{a}
	      \fmfforce{63,3}{b}

	      \fmfforce{0,23}{fb}

      \end{fmfgraph*}
	\end{fmffile}

&  \multirow{1}[4]{*}[5ex]{$ + $ }


&	\begin{fmffile}{pi2}
      \begin{fmfgraph*}(63,48)

	      \fmfpen{thin}\fmfleftn{l}{2}\fmfrightn{r}{2}
 
	      \fmfright{b,a}
	      \fmfleft{f,fb}
 	      \fmf{vanilla}{fb,v1}
	      \fmf{vanilla,tension=.5}{p1,v1,p2}
	      \fmf{vanilla}{p2,a1}
	      \fmf{vanilla}{p1,a2}
	      \fmf{vanilla}{p1,p2}
	      \fmf{vanilla}{a2,b}
	      \fmf{vanilla}{a1,a}
	      \fmf{vanilla}{p2,a1}
	      \fmf{vanilla}{a1,a}
	      \fmf{vanilla,lab.side=down,label=$a_{1}$}{p1,a2}
	      \fmf{vanilla}{a2,b}
	      \fmf{dashes}{a2,a1}	%
	      \fmfv{decor.shape=circle,decor.filled=empty,decor.size=0.06w}{a2}
	      \fmfv{decor.shape=circle,decor.filled=30,decor.size=0.06w}{a1}
   
	      \fmfforce{48,43}{p2}
	      \fmfforce{48,3}{p1}
	      \fmfforce{60,43}{a1}	%
	      \fmfforce{60,3}{a2}	%
	      \fmfforce{73,43}{a}
	      \fmfforce{73,3}{b}

	      \fmfforce{0,23}{fb}

      \end{fmfgraph*}
	\end{fmffile}

&  \multirow{1}[4]{*}[5ex]{$ + $ }

&	\begin{fmffile}{pi3}
      \begin{fmfgraph*}(63,48)

	      \fmfpen{thin}\fmfleftn{l}{2}\fmfrightn{r}{2}
 
	      \fmfright{b,a}
	      \fmfleft{f,fb}
 	      \fmf{vanilla}{fb,v1}
	      \fmf{vanilla,tension=.5}{p1,v1,p2}
	      \fmf{vanilla}{p2,a1}
	      \fmf{vanilla}{p1,a2}
	      \fmf{vanilla}{p1,p2}
	      \fmf{vanilla}{a2,b}
	      \fmf{vanilla}{a1,a}
	      \fmf{vanilla,lab.side=left,label=$a_{1}$}{p2,a1}
	      \fmf{vanilla}{a1,a}
	      \fmf{vanilla,lab.side=down,label=$a_{1}$}{p1,a2}
	      \fmf{vanilla}{a2,b}

	      \fmfv{decor.shape=circle,decor.filled=empty,decor.size=0.06w}{a2}
	      \fmfv{decor.shape=circle,decor.filled=empty,decor.size=0.06w}{a1}
   
	      \fmfforce{38,43}{p2}
	      \fmfforce{38,3}{p1}

	      \fmfforce{63,43}{a}
	      \fmfforce{63,3}{b}

	      \fmfforce{0,23}{fb}

      \end{fmfgraph*}
	\end{fmffile}

 \end{tabular}
 \end{center}
\caption{Triangle diagrams with $a_1$ -- $\pi$ transitions}
 \end{figure}

The transition $\gamma$ -- $\rho$ takes the form (see~\cite{pepan86})
\begin{equation}
\frac{e}{g_\rho}(g^{\nu \nu'}q^2 - q^{\nu} q^{\nu'})\,.
\end{equation}

Thus, one can write $B_{\rho \gamma}$ contribution in the form
\begin{equation}
B_{\gamma \rho} = 1 + \frac{s}{m_\rho^2 - s - i\sqrt{s}\Gamma_\rho(s)} = \frac{1 - i\sqrt{s}\Gamma_\rho(s)/m_\rho^2}{m_\rho^2 - s - i\sqrt{s}\Gamma_\rho(s)}m_\rho^2\, .
\end{equation}
Let us note that this expression is close to vector meson dominance model.

The term describing the transition $\gamma$ -- $\omega$ is equal of the term $\gamma$ -- $\rho$ multiplied on the factor 1/3~\cite{pepan86,ArbuzovKuraevVolkov}. The $\omega \to \pi\pi$ process was described in~\cite{pepan86}
\begin{equation}
C(m_\rho^2)\omega_\mu (p_{\pi^{+}}^\mu - p_{\pi^{-}}^\mu)\,,
\end{equation}
where $C(s) = C_1(s) + C_2(s)$.
$C_1$ describes the amplitude of transition $\omega \to \rho$ due to the difference of two quark loops. The first of them contains only $u$ quark and second contains only $d$ quark. Using the last experimental data for the decay $\omega \to \pi\pi$~\cite{PDG} we obtain $m_d - m_u \approx 3.7$ MeV. This difference allows us to describe not only the decay $\omega \to \pi\pi$, but the mass difference of charge and neutral pion and kaon (see~\cite{pepan86}) and obtain the interference $\omega$ -- $\rho$ in process $e^{+}e^{-} \to \pi^{+}\pi^{-}$ in good agreement with experimental data
\begin{equation}
C_1(s) = \frac{8(\pi\alpha_\rho)^{3/2}m_\omega^2}{3(m_\omega^2 - s - i\sqrt{s}\Gamma_\rho(s))}\frac{3}{(4\pi)^2}\log\left(\frac{m_d}{m_u}\right)^2 \, ,
\end{equation}
and $C_2$ describes the amplitude $\omega \to \gamma \to \rho$
\begin{equation}
C_2(s) = -\sqrt{\frac{\pi}{\alpha_\rho}}\frac{2\alpha s}{3(m_\omega^2 - s - i\sqrt{s}\Gamma_\rho(s))} \, .
\end{equation}

Then for the part of the amplitude with intermediate $\omega$ meson we get
\begin{equation}
B_\omega = \frac{C(s)}{3g_\rho} \frac{s}{m_\omega^2 - s - i\sqrt{s}\Gamma_\omega(s)}\, .
\end{equation}

The last part of the amplitude contains intermediate $\rho'(1450)$ meson. The transition $\gamma$ -- $\rho'$ has the form
\begin{equation}
C_{\gamma\rho'} \frac{e}{g_\rho}(g^{\nu \nu'}q^2 - q^{\nu} q^{\nu'})\, ,
\end{equation}
\begin{equation}
C_{\gamma\rho'} = -\left(\frac{\cos(\beta+\beta_0)}{\sin(2\beta_0)} + \Gamma\frac{\cos(\beta - \beta_0)}{\sin{(2\beta_0)}}\right)\, ,
\end{equation}
\begin{equation}
\Gamma = \frac{I^f_2}{\sqrt{I_2 I^{ff}_2}} = 0.54
\end{equation}

\begin{figure}[h]
\center{\includegraphics[width=0.7\linewidth]{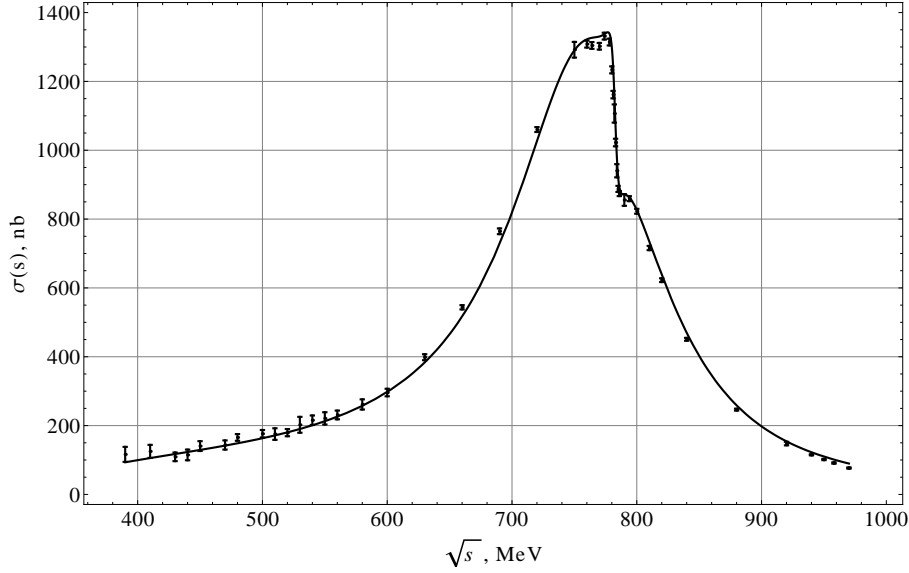}}
\label{cyl_fig}
\caption{Comparison of experimental results for $e^{+}e^{-} \to \pi^{+}\pi^{-}$ with the NJL prediction}
\end{figure}

The vertex $\rho' \pi\pi$ is proportional to
\begin{equation}
C_{\rho'\pi\pi} = -\left(\frac{\cos(\beta + \beta_0)}{\sin(2\beta_0)}g_{\rho_1} + \frac{\cos(\beta - \beta_0)}{\sin(2\beta_0)}\frac{I^f_2}{I_2}g_{\rho_2}\right) = 1.68.
\end{equation}

Unfortunatly, our model can not describe relative phase between $\rho(770)$ and $\rho'(1450)$ in $e^{+}e^{-} \to \pi\pi(\pi')$. Thus, we should get phase from $e^{+}e^{-}$ annihilation and $\tau$ decays experiments: $B_{\rho'} \to e^{i\pi}B_{\rho'}$

Thus, the $\rho'$ meson contribution reads
\begin{equation}
B_{\rho'} = e^{i\pi}\frac{C_{\gamma\rho'} C_{\rho'\pi\pi}}{g_\rho}\frac{s}{m_{\rho'}^2 - s - i\sqrt{s}\Gamma_{\rho'}(s)}\, ,
\end{equation}
where the running width $\Gamma_{\rho'}$ reads~\cite{ArbuzovKuraevVolkov}
\begin{eqnarray}
\Gamma_{\rho'}(s) &=& \Theta(2m_\pi - \sqrt{s}) \Gamma_{\rho'\to2\pi}
\\ &+& \Theta(\sqrt{s} - 2m_\pi)(\Gamma_{\rho'\to 2\pi} + \Gamma_{\rho'\to\omega\pi}\frac{\sqrt{s}-2m_\pi}{m_\omega-m_\pi})\Theta(m_\omega + m_\pi - \sqrt{s})
\nn
\\ &+& \Theta(m_{\rho'} - \sqrt{s})\Theta(\sqrt{s} - m_\omega - m_\pi) \cdot
\nn
\\ && \left(\Gamma_{\rho'\to 2\pi} + \Gamma_{\rho'\to\omega\pi}
 + (\Gamma_{\rho'} - \Gamma_{\rho'\to2\pi}  -\Gamma_{\rho'\to\omega\pi})
\nn
\frac{\sqrt{s}- m_\omega - m_\pi}{m_{\rho'} - m_\omega - m_\pi}\right)
\nn
\\ &+& \Theta(\sqrt{s} - m_{\rho'})\Gamma_{\rho'}\,.\nn
\end{eqnarray}
The values $\Gamma(\rho'\to2\pi)=22$~MeV
and $\Gamma(\rho'\to\omega\pi^0)=75$~MeV were calculated in \cite{VolkovEbertNagy}. The $\Gamma_{\rho'} = 340$ MeV was taken the value of lower boundary from PDG~\cite{PDG}.

For the total cross-section we get
\begin{equation}
\sigma(s) = \frac{\alpha^2 \pi}{12s} f_{a_1}^2(s)(1 - 4m_\pi^2/s)^{3/2} \left|B_{\rho\gamma} + B_\omega + B_{\rho'}\right|^2\,.
\end{equation}
The total cross-section is defined by $\rho$ и $\omega$ mesons, the $\rho'$ meson contributes only to the differential cross-section.

Let us note that for description of the $e^{+}e^{-} \to \pi'\pi$ not only the intermediate state with $\rho'(1450)$ but the intermediate state $\rho''(1700)$ can play the important role. However, here we take into account only rift radial excitation of $\rho$ meson. Therefore, we can pretend only to quality description of this process. The corresponding total cross-section takes the form
\begin{equation}
\sigma(s) = \frac{\alpha^2 \pi}{12s^2}\Lambda^{3/2}(s,m_{\pi'}^2,m_{\pi}^2) \left|B_{\rho\gamma}^{\pi\pi'} + B_{\rho'}^{\pi\pi'}\right|^2\,,
\end{equation}
\begin{equation*}
B_{\rho\gamma}^{\pi\pi'} = \frac{C_{\rho \pi\pi'}}{g_\rho}\left(1 + \frac{s}{m_\rho^2 - s - im_\rho\Gamma_\rho}\right) = \frac{C_{\rho \pi\pi'}}{g_\rho}\frac{1 - i\Gamma_\rho/m_\rho}{m_\rho^2 - s - im_\rho\Gamma_\rho}m_\rho^2,
\end{equation*}
\begin{equation*}
B_{\rho'}^{\pi\pi'} = e^{i\pi}\frac{C_{\gamma\rho'} C_{\rho'\pi\pi'}}{g_\rho}\frac{s}{m_{\rho'}^2 - s - im_{\rho'}\Gamma_{\rho'}},
\end{equation*}
where $\Lambda(s,m_{\pi'}^2,m_{\pi}^2) = (s - m_{\pi'}^2 - m_{\pi'}^2)^2 - 4m_{\pi'}^2m_{\pi}^2$, $m_{\pi'} = 1300$~MeV is mass of $\pi'$ meson~\cite{PDG}, $C_{\rho \pi\pi'}$ and $C_{\rho'\pi\pi'}$ is defined in a similar way as $C_{\rho'\pi\pi}$ with the use of the Lagrangian~(\ref{L2}).

\begin{figure}[h]
\center{\includegraphics[width=0.7\linewidth]{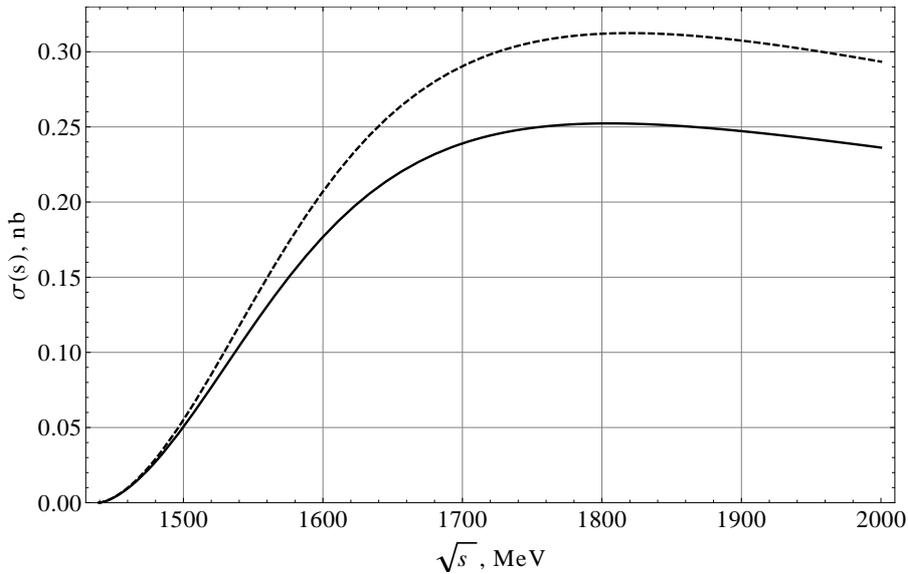}}
\label{cyl_fig}
\caption{NJL prediction for $e^{+}e^{-} \to \pi\pi'$. The solid line is the total cross-section, the dashed line is the $\rho'(1450)$ contribution only}
\end{figure}

Now for the processes $e^{+}e^{-} \to 4\pi$ we have the following experimental data: $\sigma(e^{+}e^{-}\to\pi^{+}\pi^{-}\pi^{0}\pi^{0}) \approx 10$ nb at energy $1.5$ GeV; $\sigma(e^{+}e^{-}\to\pi^{+}\pi^{-}\pi^{+}\pi^{-}) \approx 30$ nb at energy $1.5$ GeV. Therefore, we can see that our result does not contradict these data and can give a noticeable contribution to these processes. 
\section{Conclusions}
Let us note than our version of the NJL model allows us to describe not only meson production in the $e^{+}e^{-}$ processes but branching of the decay of tau lepton into mesons. Indeed, in the works~\cite{VolkovIvanovOsipov3pi,VolkovIvanovOsipovpigamma,VolkovKostunin} decays $\tau$ into $3\pi\nu$, $\pi\gamma\nu$ and $\pi\pi\nu$ were described in satisfactory agreement with experimental data.
The calculation of the last process $\tau \to \pi\pi\nu$ is very close to process $e^{+}e^{-} \to \pi^{+}\pi^{+}$ which was considered here.
In~\cite{VolkovKostunin} we obtained satisfactory agreement of both branching and differential width with experimental data.
With the help of the method use here we can obtain also a qualitative prediction for branching of the process $\tau \to \pi\pi'(1300)\nu$. This value approximately equals $0.2$\%, which does not contradict modern experimental data regarding the decays $\tau \to 4\pi\nu$. This prediction can be useful result for future experimental measurement.

\section*{Acknowledgments}
We are grateful to E.~A.~Kuraev and A.~B.~Arbuzov for useful discussions. This work was supported by RFBR grant 10-02-01295-a.

\end{document}